\documentclass[11pt]{article}
\usepackage[margin=1in]{geometry}
\usepackage{fix-cm}
\usepackage{graphicx}
\usepackage{tikz,amssymb,amsmath,amsthm}
\usepackage{newtxtext}
\usepackage{newtxmath}
\usepackage[authoryear]{natbib}
\usepackage[affil-it]{authblk}
\usepackage{lineno}
\usepackage{hyperref}
\usepackage{placeins}
\usepackage{xcolor}

\newcommand{\backsection}[2][]{\par\medskip\noindent\textbf{#1.}\ #2}

\definecolor{responseonecolor}{HTML}{B00000}
\definecolor{responsetwocolor}{HTML}{006400}
\definecolor{responsethreecolor}{HTML}{0000AA}
\definecolor{responseallcolor}{HTML}{FF69B4}

\newcommand{\rone}[1]{\textcolor{black}{#1}}
\newcommand{\rtwo}[1]{\textcolor{black}{#1}}
\newcommand{\rthree}[1]{\textcolor{black}{#1}}
\newcommand{\ra}[1]{\textcolor{black}{#1}}

\hypersetup{ colorlinks = true, urlcolor = blue, citecolor = black, }

\newcommand{\RomanNumeralCaps}[1]
\linenumbers
\graphicspath{{./}{figures/}}
\title{\ra{On the Poisson-Source Basis of Logarithmic Wall-Pressure-Variance Growth}}

\author[1]{J. M. O. Massey\thanks{Email: \href{mailto:masseyj@stanford.edu}{masseyj@stanford.edu}}}
\author[2]{J. C. Klewicki}
\author[1]{B. J. McKeon}
\affil[1]{Center for Turbulence Research, Stanford University, Stanford, CA 94305, USA}
\affil[2]{Department of Mechanical Engineering, University of Melbourne, Parkville, Victoria 3010, Australia}
\date{}

\begin{document}
\maketitle

\begin{abstract}
  In high-Reynolds-number wall-bounded flows, the inner-scaled wall-pressure variance \ra{is often represented as a} logarithmic increase with frictional Reynolds number.
  We consider the two sources of the incompressible pressure--Poisson equation: a linear (rapid) term linked to mean shear and a nonlinear (slow) term composed of quadratic velocity fluctuations. This paper establishes a link between the sources and the coefficients in \rtwo{a logarithmic} inner-scaled variance \rtwo{representation}. \rone{To leading order} we \rone{posit that} the \ra{linear source provides a Reynolds-number-independent offset, while the nonlinear source contributes the logarithmic coefficient}. The illustrative dataset is direct numerical simulation (DNS) at \rtwo{frictional Reynolds number} $\delta^+\approx 550$, although the principal contribution is the establishment of a mechanistic link to well-known high-$\delta^+$ scalings of wall-bounded turbulence.
  Through consideration of the sources and the integral solution method of the Poisson equation, we find that the linear source contribution sits predominantly in the buffer layer and maps to the near-wall cycle. \rone{To leading order}, this contribution becomes $\delta^+$ invariant under inner scaling, thus contributing an offset in the \rtwo{logarithmic representation}.
  The interfacial regions between uniform momentum zones characteristic of the inertial layer (vortical fissures) spatially localise strain and vorticity contributions and contain an increasingly large proportion of the strain and vorticity. We show that fissures act as a compact carrier for the source terms, with the nonlinear term especially prominent in these regions. By considering the inertial layer statistics, we link the changing nonlinear contribution to \rtwo{the} $\ln \delta^+$ \rtwo{growth}, in agreement with previous empirical observations.
\end{abstract}

\section{Introduction}

  Wall-pressure fluctuations in turbulent wall-bounded flows underpin acoustic noise, structural fatigue, and the performance of flow-control devices. In incompressible turbulence, the pressure satisfies a Poisson equation whose right-hand side contains two source terms: a linear (rapid) term coupling the mean shear to the fluctuating velocity field and a nonlinear (slow) term from quadratic velocity--velocity interactions; a viscous (Stokes) term arises from the Neumann boundary condition enforcing no slip at the wall. As the friction Reynolds number, $\delta^+\equiv \delta u_\tau/\nu$, increases, the viscous contribution becomes progressively less important and the linear and nonlinear mechanisms dominate \citep{kim_structure_1989,jimenez2008turbulent,gerolymos_wall_2013}. Here, $\delta$ is the boundary-layer thickness and half-height in channel, $u_\tau$ is the friction velocity, and $\nu$ is the kinematic viscosity.

  \begin{figure}
    \centering
    \includegraphics{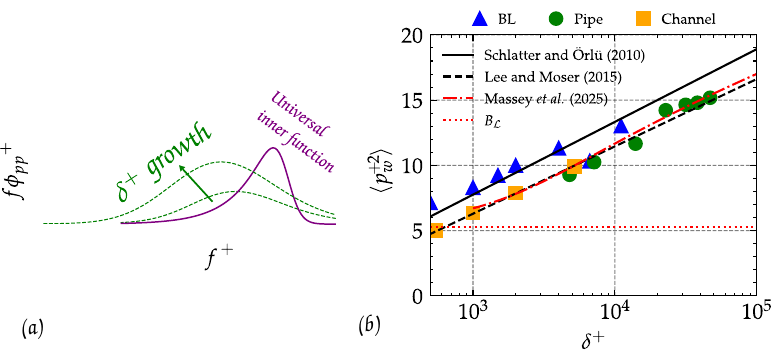} 
    \caption{(a) Schematic of the wall-pressure spectra from \citet{massey_eddy_2025}, showing a universal---$\delta^+$-independent---inner function and a $\delta^+$-dependent outer function responsible for modelling the contribution from outer-scaled motions. \rthree{The axes are omitted to illustrate this is a schematic, not a data plot}. (b) $\langle p_w^{+2}\rangle$ for various DNS and experiments, showing the $\ln\delta^+$ growth. Blue triangles represent boundary layer data from \citet{fritsch_pressure_2020,fritsch_fluctuating_2022}, green circles represent pipe data from \citet{dacome_scaling_2025}, and orange squares represent channel data from \citet{lee_direct_2015}. The red dotted line is the variance from the universal inner function, showing that it contributes an $O(1)$ offset, $B_{\mathcal L}$, when the inner function is fully developed.}
    \label{fig:inner- outer-function schematic}
  \end{figure}

  Experiments and simulations have mapped the spectral and spatial origins of wall pressure: High frequencies are associated with the near-wall cycle, intermediate frequencies with the logarithmic region, and low frequencies with the outer flow \citep{farabee_spectral_1991,gibeau_low-_2021,dacome_scaling_2025,deshpande_source_2025}. DNS at moderate $\delta^+$ confirms that buffer layer eddies contribute strongly to both linear and nonlinear pressure, with viscous effects confined to the smallest scales \citep{kim_structure_1989,choi_space-time_1990,chang_relationship_1999,hu_wall_2006,lee_direct_2015,anantharamu_analysis_2020,yang_wavenumber-frequency_2022}. The wall-normal structure of the sources has been examined directly \citep{kim_structure_1989,anantharamu_analysis_2020}. More broadly, wall-flow theory points to inner (viscous-scaled) and outer (boundary layer--scaled) regions \citep{bradshaw_inactive_1967,panton_correlation_2017,deshpande_active_2025}, and that pressure can be modelled as contributions from these separate regions. Figure~\ref{fig:inner- outer-function schematic}(a) graphically depicts the procedure proposed by \citet{massey_eddy_2025}. \rthree{Figure~\ref{fig:inner- outer-function schematic}(a) describes the spectral structure of the wall-pressure itself; no attribution of specific spectral features to $\mathcal{L}$ or $\mathcal{Q}$ is implied.}

  A key empirical observation, from high-$\delta^+$ boundary layers and pipes \citep{klewicki_statistical_2008,fritsch_pressure_2020,fritsch_fluctuating_2022,dacome_scaling_2025} as well as from DNS flows \citep{hoyas_scaling_2006,schlatter_assessment_2010,lee_direct_2015,pirozzoli_pressure_2025}, is the logarithmic growth of the inner-scaled wall-pressure variance, $\langle p_w^{+2}\rangle\propto\ln \delta^+$, where $p_w^+\equiv p_w/(\rho u_\tau^2)\equiv p_w/\tau_w$ (Figure~\ref{fig:inner- outer-function schematic}(b)). Matched-asymptotic descriptions reproduce this growth \citep{panton_correlation_2017}, which is consistent with a progressive broadening of the low-wavenumber portion of the wall-pressure spectra. This finding suggests that the responsible physical processes are tied to the broadening of the log layer connecting the buffer and outer layer.

  With homogeneous Neumann boundary conditions, the Green's function of the Poisson operator decays with wall-normal distance $y$ and, in Fourier space, with the streamwise--spanwise wavenumber magnitude $k=\sqrt{k_x^2+k_z^2}$. \rthree{The ring-averaged representation is used only to expose the radial Green-function attenuation; the underlying wall-pressure and source fields remain anisotropic in $(k_x,k_z)$.} \rtwo{Throughout, we use $(x,y,z)$ for the streamwise, wall-normal, and spanwise coordinates, respectively. When tensor notation is convenient, repeated Roman indices refer to components in this same Cartesian frame.} Approximately, the Green's function decays proportionally to ${\rm e}^{-k y}$~\citep{anantharamu_analysis_2020}; short-wavelength pressure fluctuations are supplied predominantly by near-wall source terms, whereas long-wavelength pressure is supplied by sources farther from the wall. This provides a bridge between wall-pressure statistics and the wall-normal distribution of the pressure sources. This nonlocal mapping motivates a relation between polar spectral content at the wall and the locations and types of Poisson sources across $y$.

  A growing body of theory, DNS, and experimental work indicates that the inertial region organises into large uniform-momentum zones (UMZs) separated by narrow, locally sheet-like fissures bearing concentrated spanwise vorticity~\citep{meinhart_existence_1995,priyadarshana_statistical_2007,de_silva_interfaces_2017,heisel_self-similar_2022} and in which enstrophy and strain are both large and strongly correlated very near the wall, vorticity is 2-D and space filling; it then undergoes stretching and tilting through the viscous and buffer layers. Vorticity in the inertial layer starts aligning in the streamwise direction under the action of mean shear and concentrates within thin vortical fissures. In this UMZ-vortical-fissure picture, the fissure thickness, when scaled by outer length, decreases approximately as $1/\sqrt{\delta^+}$, while the streamwise velocity jump across a fissure is $O(u_\tau)$; both are central empirical inputs in minimal models that reproduce turbulent statistics at high $\delta^+$~\citep{bautista_uniform_2019}. Theoretical analyses of the mean momentum equation further support the view that with increasing $\delta^+$ an increasing fraction of the vorticity outside the buffer region is confined within such narrow fissures that disperse across a hierarchy of wall-normal layers~\citep{klewicki_description_2013,morrill-winter_influences_2013}.

  At high $\delta^+$ incompressible wall turbulence, the wall-pressure variance \ra{is often represented by}
  \begin{equation}
    \langle p_w^{+2}\rangle = B_\mathcal{L} + A_\mathcal{Q}\,\ln \delta^+ .
    \label{eq:pressure_variance_law}
  \end{equation}
  \rone{Empirical fits give, for example, $\langle p_w^{+2}\rangle = 2.42\ln\delta^+ - 8.96$ for boundary-layer data \citep{schlatter_assessment_2010} and $\langle p_w^{+2}\rangle = 2.24\ln\delta^+ - 9.18$ for channel-flow data \citep{lee_direct_2015}.}
  \rtwo{We stress that equation~\eqref{eq:pressure_variance_law} is adopted here as a working asymptotic hypothesis, motivated by the currently available wall-pressure data, rather than as an uncontested starting point. More broadly, the existence and universality of logarithmic overlap laws for fluctuation statistics in wall turbulence remain debated, and alternative bounded or quarter-power forms have been proposed from both asymptotic and symmetry-based perspectives \citep{chen_reynolds_2023,oberlack_turbulence_2022,hoyas2022wall}. In particular, \citet{chen_reynolds_2023} argue for a bounded outer form for several wall-bounded fluctuation quantities, including the root-mean-square pressure, while \citet{nagib_utilizing_2024} revisit the logarithmic-versus-quarter-power interpretation for normal stresses using indicator functions applied to computational data. The present paper does not attempt to resolve that broader debate. Instead, conditional on the empirical wall-pressure form in equation~\eqref{eq:pressure_variance_law}, we ask what Poisson-source mechanisms would be consistent with such behaviour and what that implies for the coefficients $B_\mathcal{L}$ and $A_\mathcal{Q}$.}

  In this paper, we \rone{provide evidence to support the} link \rone{between} equation~\eqref{eq:pressure_variance_law} \rone{and} the processes governing the behaviour of the Poisson sources themselves. Using channel DNS at $\delta^+\approx 550$, we show that the elliptic attenuation of the Green's function maps contributions from the source at $yk \approx O(1)$. The root mean square (r.m.s.) of the source terms shows that the linear source peaks in the buffer layer, while the nonlinear source is made up of a balance of strain and enstrophy peaking in the inertial layer. Next, we show that the linear source--to leading order--is tied to the near-wall cycle and asymptotes with $\delta^+$ to an $O(1)$ offset to the logarithmic $\langle p_w^{+2}\rangle$ growth, arising as $B_\mathcal{L}$ in equation~\eqref{eq:pressure_variance_law}. Finally, we show that inertial layer fissures act as compact carriers of the nonlinear source that contributes to $A_\mathcal{Q}\ln \delta^+$. This connection of the nonlinear source to inertial layer processes, and thus statistics, provides a mechanistic link between the nonlinear source and the $\ln\delta^+$ growth of $\langle p_w^{+2}\rangle$. \ra{Because the detailed source decomposition is available only at $\delta^+\approx 550$, these proposed $\delta^+$ trends are necessarily inferential: they combine the single-Reynolds-number DNS with prior high-Reynolds-number evidence and scaling arguments, and therefore should be read as hypotheses about the asymptotic regime rather than as a direct verification of generality. Nevertheless, we regard this case as adequate for mechanistic illustration, the viscous contribution is already subdominant, and the source-localisation and wall-mapping mechanisms are directly visible in the decomposed fields.} Finally, we use our simplifying assumptions to predict $A_\mathcal{Q}\approx 2$, in good agreement with previous empirical observations.
  
\section{Components of the pressure Poisson equation}\label{sec:poisson_components}
  \begin{figure}
    \centering
    \includegraphics[width=0.9\linewidth]{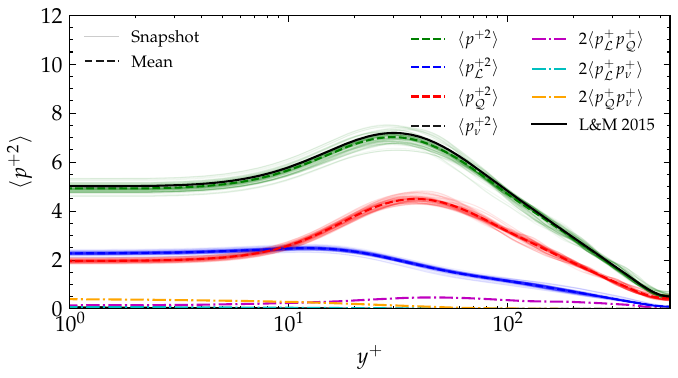}
    \caption{Instantaneous and mean r.m.s. profiles of the pressure across a half-channel width at $\delta^+\approx 550$ and a comparison to the r.m.s. reported by \citet{lee_direct_2015}. The linear, nonlinear, and viscous contributions to the variance as well as the cross terms are shown. \rthree{Light curves indicate instantaneous profiles, while the darker/dashed curves indicate the corresponding mean profiles.}}
  \label{fig:pressure_profile}
  \end{figure}

  Following \citet{mansour_reynolds-stress_1988} and \citet{kim_structure_1989}, and using the Einstein summation convention \rtwo{where Roman indices are just component indices in that same Cartesian frame}, the fluctuating pressure satisfies
  \begin{equation}
    \partial_i\partial_i p
    = \underbrace{-\,2\,\frac{{\rm d}U}{{\rm d}y}\,\partial_x v}_{\text{linear }(\mathcal{L})}
    \;+\; \underbrace{-\,\partial_j\partial_k \left(u_j u_k\right)}_{\text{nonlinear }(\mathcal{Q})}
    \;+\;\overline{\,\partial_j\partial_k \left(u_j u_k\right)}\text{ ,}
    \label{eq:pp}
  \end{equation}
  where $\tilde{u}_i = U_i + u_i$. In the resolvent analysis framework \citep{mckeon_critical-layer_2010}, $\mathcal Q$ is the dilatational part of the nonlinear forcing to the linearised Navier--Stokes operator, namely the gradient of the Reynolds stress tensor \citep{rosenberg_efficient_2019,blanco_linear_2024}. Rewritten in vorticity/strain form by considering the symmetric and antisymmetric parts of the Reynolds stress gradient tensor \citep{bradshaw_note_1981}, the equation reads
  \begin{equation}
    \partial_i\partial_i p
    = \underbrace{2\,\Omega_z\,\partial_x v}_{\mathcal{L}}
    \;+\;  \underbrace{\Big(\overbrace{\tfrac12\,\omega_i\omega_i}^{\mathcal{T}}
    \;-\; \overbrace{S_{ij}S_{ij}}^{\mathcal{S}}\Big)
    \;-\;\overline{\Big(\tfrac12\,\omega_i\omega_i - S_{ij}S_{ij}\Big)}}_{\mathcal{Q}}\text{ ,}
    \label{eq:pp_vorticity}
  \end{equation}
  where
  \begin{align}
    S_{ij} &= \tfrac12(\partial_i u_j + \partial_j u_i),\qquad
    \omega_i = \varepsilon_{ijk}\,\partial_j u_k,\qquad
    \Omega_z = -\,\frac{{\rm d}U}{{\rm d}y}\text{ ,}
  \end{align}
  with $\mathcal Q = \mathcal T - \mathcal S$.

  \noindent We define the pressure components by
  \begin{subequations}\label{eq:pressure_components}
    \begin{align}
      \partial_i\partial_i p_{\mathcal{L}} &= \mathcal{L}, &\qquad (\partial_y p_{\mathcal{L}})\big|_{y=\pm 1} &= 0,\\
      \partial_i\partial_i p_{\mathcal{Q}} &= \mathcal{Q}, &\qquad (\partial_y p_{\mathcal{Q}})\big|_{y=\pm 1} &= 0,\\
      \partial_i\partial_i p_{\nu} &= 0, &\qquad (\partial_y p_{\nu})\big|_{y=\pm 1} &= Re^{-1}\,\partial_j\partial_j v \text{ ,}
    \end{align}
  \end{subequations}
  with $Re \equiv \delta U_e / \nu$, and $U_e$ is the centreline/edge velocity in the channel/boundary layer. It follows that
  \begin{equation}
    p = p_{\mathcal{L}} + p_{\mathcal{Q}} + p_{\nu}\text{ ,}
  \end{equation}
  where $p_{\mathcal{L}}$, $p_{\mathcal{Q}}$, and $p_{\nu}$ are the linear, nonlinear, and viscous components of the pressure, respectively.

  Because the analysis below is restricted to $k=\sqrt{k_x^2+k_z^2}>0$, the wall-parallel mean $\overline{\mathcal Q}(y)$ contributes only to the $(k_x,k_z)=(0,0)$ mode and therefore does not appear explicitly in the Fourier-space expressions below.

  Let $\widehat{\cdot}$ denote the $(k_x,k_z)$ Fourier transform and $k=\sqrt{k_x^2+k_z^2}>0$. In a channel $y\in[-1,1]$ with homogeneous Neumann boundary conditions, the solution of equation~\eqref{eq:pp} for the linear and nonlinear sources $\mathcal{M}\in\{\mathcal{L},\mathcal{Q}\}$ admits the integral representation
  \begin{subequations}\label{eq:poisson_integral_fullfield}
  \begin{align}
    \widehat{p}_{\mathcal{M}}(k,y;\delta^+)
    = \int_{-1}^{1} g(k;\,y,y')\,\widehat{\mathcal{M}}(y',k;\delta^+)\,{\rm d}y'\text{ ,} \\
    g(k;\,y,y')
    = -\frac{\cosh\!\big(k(1-y_>)\big)\,\cosh\!\big(k(1+y_<)\big)}{k\,\sinh(2k)} \text{ ,}
  \end{align}
  \end{subequations}
  with $y_< = \min(y,y')$ and $y_> = \max(y,y')$, where $y'\in[-1,1]$ is the source-point integration variable. The viscous component $p_{\nu}$ is obtained through enforcement of no slip via the Neumann boundary condition.

  Evaluating~equation \eqref{eq:poisson_integral_fullfield} at either wall yields a convolution between the sources and a single-wall kernel. For the bottom wall ($y=-1$),
  \begin{subequations}\label{eq:poisson_integral_walls}
    \begin{align}
    \widehat{p}_{\mathcal{M},w^-}(k;\delta^+)
    &= \int_{-1}^{1} G^{-}(k,y')\,\widehat{\mathcal{M}}(y',k;\delta^+)\,{\rm d}y',
    \;\;
    G^{-}(k,y') = -\,\frac{1}{k}\,\frac{\cosh\!\big(k(1-y')\big)}{\sinh(2k)},
    \label{eq:greens_bottom}\\
    \text{and top wall ($y=+1$),}\nonumber\\
    \widehat{p}_{\mathcal{M},w^+}(k;\delta^+)
    &= \int_{-1}^{1} G^{+}(k,y')\,\widehat{\mathcal{M}}(y',k;\delta^+)\,{\rm d}y',
    \;\;
    G^{+}(k,y') = -\,\frac{1}{k}\,\frac{\cosh\!\big(k(1+y')\big)}{\sinh(2k)}.
    \label{eq:greens_top}
  \end{align}
  \end{subequations}
  \rone{For later convenience, we define the inner-scaled distances from the bottom and top walls as
  \begin{equation}
    y_-^+ \equiv \delta^+(1+y), \qquad y_+^+ \equiv \delta^+(1-y),
  \end{equation}
  or, equivalently, $y_\pm^+ \equiv \delta^+(1\mp y)$.}

  Writing the hyperbolic functions in exponential form makes the wall-normal attenuation explicit,
  \begin{equation}
    G^{-}(k,y')
    = -\frac{1}{k}
    \frac{{\rm e}^{-k(1+y')}+{\rm e}^{-k(3-y')}}{1-{\rm e}^{-4k}}
    \;\;\xrightarrow[k\gg1]{}\;\;
    -\frac{1}{k}{\rm e}^{-k(1+y')} ,
    \label{eq:Greens k gtr 1}
  \end{equation}
  so that each Fourier mode decays exponentially with the wall-normal distance from the source to the wall. Equivalently, the full-field kernel in~equation \eqref{eq:poisson_integral_fullfield} has the leading-order form
  \begin{equation}
    g(k;y,y') \; \sim \; -\frac{1}{2k}{\rm e}^{-k|y-y'|},
  \end{equation}
  demonstrating the expected elliptic attenuation: Contributions at wavenumber~$k$ are exponentially suppressed away from their source with characteristic scale~$1/k$. Below, we focus on the wall pressure and its decomposition,
  \begin{equation}
    p_w \;=\; (p_{\mathcal{L}})_w + (p_{\mathcal{Q}})_w + (p_{\nu})_w,
  \end{equation}
  with each component obtained by substituting $\widehat{\mathcal{M}}$ into equation~\eqref{eq:poisson_integral_walls}, and we drop the $\cdot_w$ subscript for brevity.
  In the bottom-wall attribution maps below, $y$ denotes source distance from that wall, so $G(k,y)$ denotes the corresponding single-wall kernel; for $k\gg1$, $G(k,y)\sim-k^{-1}{\rm e}^{-ky}$.

  The dataset used derives from a previously published channel-flow DNS at $\delta^+\approx550$ \citep{huang_spatio-temporal_2025} that employs the method of \citet{flores_effect_2006}. Figure~\ref{fig:pressure_profile} shows a comparison with results from \citet{lee_direct_2015} (as reported in \citet{panton_correlation_2017}) and confirms that we match $\langle p^2\rangle^+$ at comparable $\delta^+$. \ra{Figure~\ref{fig:pressure_profile} shows that this case lies close to the DNS of \citet{lee_direct_2015} and that the viscous contribution is already small compared with the linear and nonlinear terms, as noted by \citet{kim_structure_1989} and \citet{jimenez2008turbulent}. We use $\delta^+\approx 550$ as an illustrative dataset: it is not sufficient to settle asymptotic scaling, but it is sufficient to expose the source localisation and wall-mapping mechanisms on which the present interpretation rests.} The nonlinear pressure variance is comparable to or larger than the linear contribution everywhere except very near the wall, with the mean-square nonlinear source peaking near $y^+\approx20$, in line with previous work \citep{kim_structure_1989,panton_correlation_2017,yang_wavenumber-frequency_2022}. \rone{The cross terms contributing to the variance are also shown. Although \citet{yang_wavenumber-frequency_2022} show that such cross terms can contribute significantly, in the present case they remain smaller than the dominant linear and nonlinear self-contributions. We therefore focus below on the mechanisms associated with $\mathcal{L}$ and $\mathcal Q$.}

\section{Source-term structure}\label{sec:source_structure}

  \begin{figure}
    \centering
    \includegraphics[width=\linewidth]{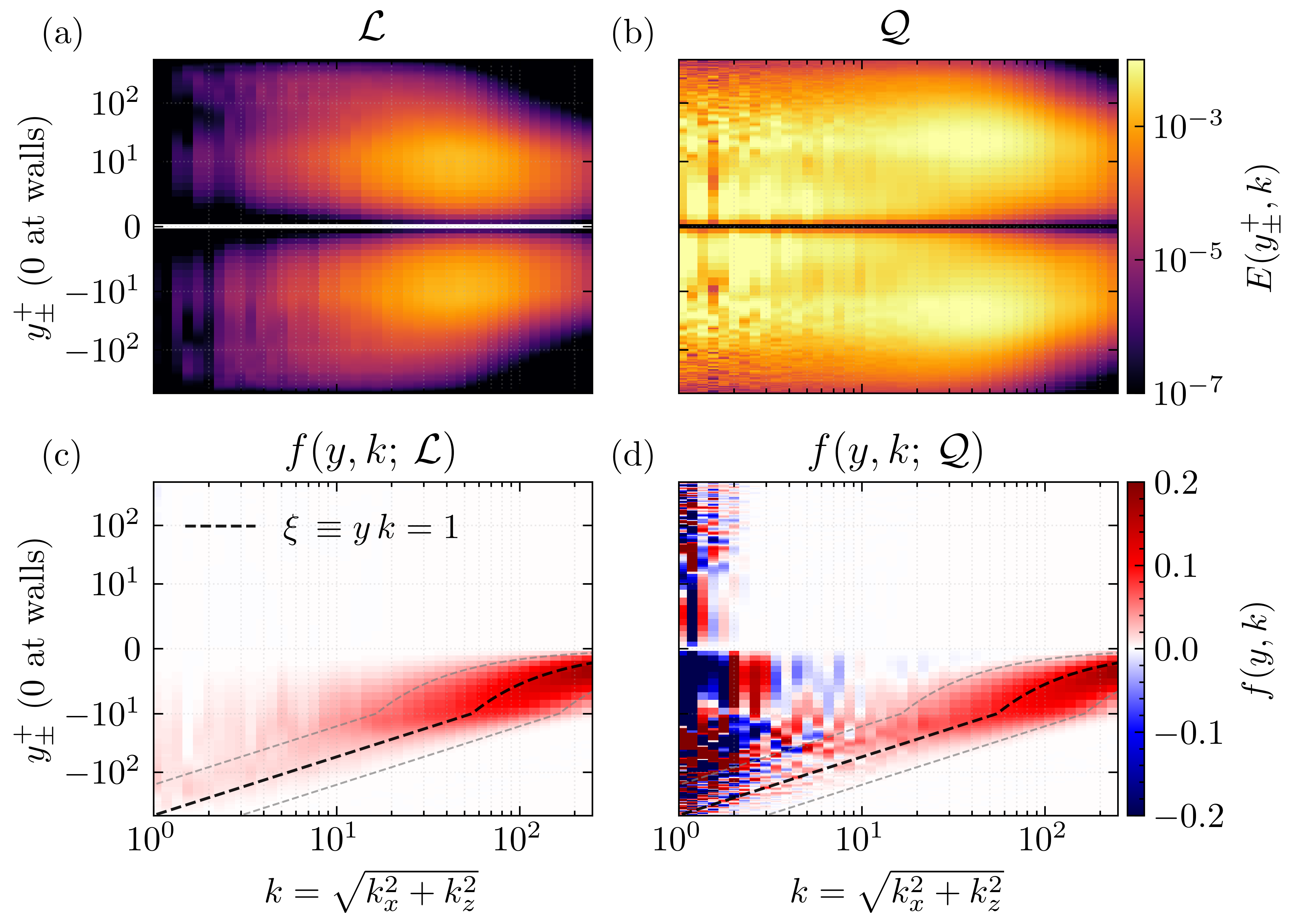}
    \caption{Ring-averaged $y^+$--$k$ spectra of (a) linear $\mathcal{L}$ and (b) nonlinear $\mathcal{Q}$ sources, as well as the corresponding wall-pressure attribution maps to the bottom wall for (c) linear pressure and  (d) nonlinear pressure. (c,d) The black dashed line indicates $k\,y=1$, and the \rthree{two gray lines indicate $\xi=0.3$ and $\xi=3$}. The kink in the lines is due to the change from log to linear scaling at $y^+=10$ toward the channel wall.}
  \label{fig:source_spectra}
  \end{figure}

  The structure of the source terms is illustrated in Figure~\ref{fig:source_spectra}. Figure~\ref{fig:source_spectra}(a) shows that the $\mathcal{L}$ energy is concentrated near $y^+\approx10$ and at relatively high $k$, whereas Figure~\ref{fig:source_spectra}(b) shows that $\mathcal{Q}$ peaks further out ($y^+\approx20$) and extends to the highest resolved $k$. As equation~\eqref{eq:pp_vorticity} shows, $\mathcal{Q}$ is the difference of two large, correlated quadratic forms. Strong cancellations occur when it is mapped onto the wall \rone{which makes it susceptible to sign changes and partial cancellation. When mapped to the wall, this yields the positive and negative attribution regions seen in figure~\ref{fig:source_spectra}(d).}

  \subsection{Variance attribution}

    Figures~\ref{fig:source_spectra}(c,d) highlight how the source terms are mapped to the wall-pressure fluctuations by showing the instantaneous wall-pressure attribution maps for $\mathcal{L}$ and $\mathcal{Q}$. \rone{Here, consistent with the bottom-wall maps, $y$ denotes source distance from the bottom wall. The attribution is defined as the signed fractional contribution of sources at distance $y$ to the wall-pressure mode at wavenumber $k$. We define the attribution function $f(y,k)$ as}
    \rone{\begin{equation}
      f(y,k) = \frac{ \Re\{ G(k,y)\,\widehat{\mathcal{M}}(y,k)\,\widehat{p}^*(k)\}}{|\widehat{p}(k)|^2} \text{ ,}
    \label{eq:attribution}
    \end{equation}
    where $\Re$ denotes the real part. For each $k$, $\int_{0}^{2} f(y,k)\,{\rm d}y = 1$; positive values indicate contributions that are in phase with the wall-pressure mode, whereas negative values indicate cancellation. Figure~\ref{fig:source_spectra}(c) shows that the attribution associated with $\mathcal{L}$ is everywhere positive. For $k \lesssim 10$ it is distributed over a range of wall-normal distances, whereas with increasing $k$ it becomes concentrated over a narrower band. The attribution associated with $\mathcal{Q}$ in figure~\ref{fig:source_spectra}(d) is similar for $k\gtrsim10$. For $k\lesssim10$, however, the nonlinear map develops alternating positive and negative regions, indicating that the wall-pressure mode is assembled through partial cancellation among source layers, including appreciable contributions from both sides of the channel centreline.}

  \subsection{Elliptic attenuation}\label{sec:elliptic attenuation}

    Heuristically, the elliptic attenuation reflects the balance between the algebraic spectral decay of the source and the exponential decay of the Green's kernel with wall-normal distance. For a wall-parallel wavenumber magnitude $k$ at height $y$, the wall-weighted amplitude scales as
    \begin{equation}
      |G(k,y)\,\widehat{\mathcal{M}}(y,k)| \;\sim\; \frac{{\rm e}^{-k y}}{k}\times\big[\text{algebraic source decay in }k\big],
    \end{equation}
    where the factor ${\rm e}^{-ky}$ comes from the Green's kernel for $k\gg1$ \eqref{eq:Greens k gtr 1}, so that contributions from a layer at $y$ are rapidly attenuated once $k\,y\gg1$. Accordingly, we use this simplifying assumption to define the effective envelope of wall-source communication by a fixed iso-attenuation criterion on the kernel; for instance,
    \begin{equation}
      {\rm e}^{-k y} = {\rm e}^{-1}\quad\Longleftrightarrow\quad \xi \equiv k\,y = 1,
      \label{eq:elliptic_line}
    \end{equation}
    where $\xi=\mathcal{O}(1)$ denotes one ${\rm e}$-fold of attenuation. For $\xi\ll1$, the kernel hardly filters (and the communication to the wall is then controlled by the source spectrum), whereas for $\xi\gg1$, the exponential damping dominates and suppresses the contribution. The line $\xi=1$ therefore serves as a practical elliptic attenuation boundary in $(k,y)$-space.

    Figures~\ref{fig:source_spectra}(c,d) plot the $\xi = 1$ line as an approximate envelope: high-$k$ contributions from a given layer lie predominantly on or slightly above this curve, while the low-$k$ contributions ($k \lesssim 10$) extend more broadly, with $\xi=1$ providing an upper bound on their influence. This low-$k$ behaviour is influenced by channel-specific effects, in particular the cancellation of contributions across the centreline and the relative (to ZPG BL flows) suppression of the largest scales. To keep the theoretical picture consistent with single-wall geometries, we therefore focus, for the present envelope argument, on the region $k>10$, and reiterate that the $e^{-ky}$ decay used in \eqref{eq:Greens k gtr 1} is the large-$k$ ($k\gg 1$) asymptotic form of the Green's kernel. Figure~\ref{fig:source_spectra}(c,d) also plots lines at $\xi=0.3$ and $\xi=3$ to give a sense of the $O(1)$ range of the elliptic attenuation.

  \subsection{Positive and negative $\mathcal{Q}$}

    It is useful to distinguish the local enstrophy--strain imbalance,
    \begin{equation}
      \mathcal{B}\equiv \mathcal{T}-\mathcal{S},
    \end{equation}
    from the nonlinear Poisson source,
    \begin{equation}
      \mathcal{Q}=\mathcal{B}-\overline{\mathcal{B}}.
    \end{equation}
    Regions with $\mathcal{B}>0$ are locally enstrophy-dominated, whereas regions with $\mathcal{B}<0$ are locally strain-dominated. By contrast, the sign of $\mathcal{Q}$ indicates whether the local enstrophy--strain imbalance lies above or below its mean, not simply whether $\mathcal{T}$ is larger or smaller than $\mathcal{S}$. Thus, $\mathcal{Q}>0$ and $\mathcal{Q}<0$ identify positive and negative nonlinear source events, respectively.

    To isolate these two classes of events, we define
    \begin{equation}
      \mathcal{Q}^\oplus=\max(\mathcal{Q},0),
      \qquad
      \mathcal{Q}^\ominus=\max(-\mathcal{Q},0),
      \label{eq:strain enstrophy domination}
    \end{equation}
    and examine their r.m.s. profiles separately.

  \subsection{Root mean square of the source components}
    \begin{figure}
      \centering
      \includegraphics[width=\linewidth]{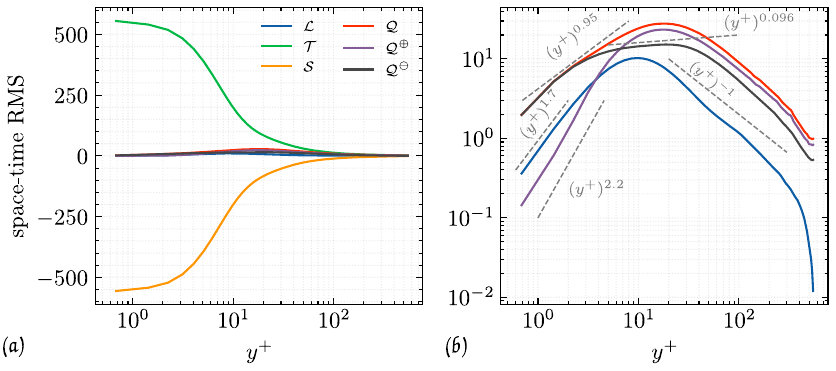}
      \caption{Profiles of the r.m.s. of the Poisson source components versus $y^+$. (a) The individual components, $\mathcal{L}$, $\mathcal{T}$, $-\mathcal{S}$, and $\mathcal{Q}$, and the sign-split parts, $\mathcal{Q}^\oplus$ and $\mathcal{Q}^\ominus$ (Eqs.~\eqref{eq:pp},\eqref{eq:pp_vorticity}, and \eqref{eq:strain enstrophy domination}). (b) Log--log view of selected curves.}
      \label{fig:rms_all_terms}
    \end{figure}

    Profiles of the r.m.s. of the Poisson-source terms versus $y^+$ quantify the available source amplitude prior to cancellation by sign and by Green-function weighting. Figure~\ref{fig:rms_all_terms}(a) shows that $\mathcal{T}$ and $\mathcal{S}$ are individually much larger than $\mathcal{L}$ and $\mathcal{Q}$, so that the nonlinear source arises from a strong cancellation between large enstrophy and strain contributions. Figure~\ref{fig:rms_all_terms}(b) shows the selected r.m.s. profiles on logarithmic axes. The linear source $\mathcal{L}$ peaks at $y^+\approx 10$, whereas $\mathcal{Q}$ peaks farther from the wall, at $y^+\approx 15$--$20$, and remains larger than $\mathcal{L}$ over most of the domain. Over the inertial layer, both $\mathcal{L}$ and $\mathcal{Q}$ decay approximately as $y^{+-1}$.

    The $\mathcal{Q}^\oplus/\mathcal{Q}^\ominus$ split isolates positive and negative nonlinear source events, rather than enstrophy- and strain-dominated events. Very near the wall, $\mathcal{Q}^\ominus_{\mathrm{rms}} > \mathcal{Q}^\oplus_{\mathrm{rms}}$. Beyond $y^+\approx 7$, $\mathcal{Q}^\oplus_{\mathrm{rms}}$ rises rapidly and becomes comparable to, and near the peak slightly larger than, $\mathcal{Q}^\ominus_{\mathrm{rms}}$; farther from the wall the two remain of the same order.
\section{Direct source density of the linear contribution}\label{sec:L_source}

\begin{figure}
    \centering
    \includegraphics[width=\linewidth]{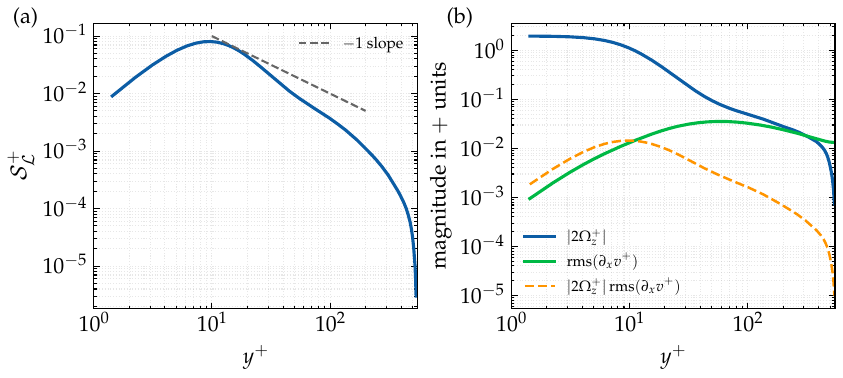}
    \caption{
    Linear-source variance-density decomposition at $\delta^+\approx550$.
    (a) Direct signed wall-normal source density of the linear pressure variance, $\mathcal S_{\mathcal L}^+(y^+)$, defined by \eqref{eq:variance_source_density}; the solid curve shows the positive part and the dotted curve shows the magnitude of the negative part.
    (b) Inner-scaled magnitudes of the two local factors entering $\mathcal L$: $|2\Omega_z^+|$, ${\rm r.m.s.}(\partial_{x^+}v^+)$, and their pointwise product. Comparing (a) and (b) distinguishes local source amplitude from the Green-function-weighted, phase-aware contribution to wall-pressure variance.
    }
    \label{fig:dvdx_linear_source_attribution}
\end{figure}

The $\delta^+$-invariant inner function of the wall-pressure spectrum modelled by \citet{massey_eddy_2025} is defined only once $\delta^+$ is sufficiently large. In practice, the proposed $B_{\mathcal L}$ offset in equation~\eqref{eq:pressure_variance_law} is the asymptotic value obtained when that inner contribution is integrated to $\langle p^{+2}\rangle$. The question is therefore whether the linear self-contribution can generate a logarithmic dependence as the wall-normal domain expands with $\delta^+$.

The r.m.s. profiles discussed above quantify the amplitude available before wall mapping. Figure~\ref{fig:dvdx_linear_source_attribution}(b) shows that the local product $|2\Omega_z^+|\,{\rm r.m.s.}(\partial_{x^+}v^+)$ is weighted toward the near-wall region. This is suggestive, but not sufficient: it does not include the Green-function weighting, wall-normal correlations, or cancellations between source layers.

A direct test is obtained from the wall-normal source density of the variance. For a source component $\mathcal M\in\{\mathcal L,\mathcal Q\}$, multiplying equation~\eqref{eq:poisson_integral_walls} by $\widehat p_{\mathcal M}^{\,*}$, summing over wall-parallel Fourier modes, and averaging gives
\begin{equation}
    \langle p_{\mathcal M}^{2}\rangle
    =
    \int_{-1}^{1}\mathcal S_{\mathcal M}(y)\,dy,
    \label{eq:variance_source_density_integral}
\end{equation}
with the signed wall-normal source density
\begin{equation}
    \mathcal S_{\mathcal M}(y)
    =
    \sum_{\boldsymbol{k}}
    \Re\left\{
    G_w(k,y)
    \left\langle
    \widehat{\mathcal M}(\boldsymbol{k},y)\,
    \widehat p_{\mathcal M}^{\,*}(\boldsymbol{k})
    \right\rangle
    \right\}.
    \label{eq:variance_source_density}
\end{equation}
Equivalently,
\begin{equation}
    \mathcal S_{\mathcal M}(y)
    =
    \int_{-1}^{1}
    \sum_{\boldsymbol{k}}
    G_w(k,y)G_w(k,y')
    \Re\left\{
    \left\langle
    \widehat{\mathcal M}(\boldsymbol{k},y)
    \widehat{\mathcal M}^{\,*}(\boldsymbol{k},y')
    \right\rangle
    \right\}
    dy'.
    \label{eq:variance_source_density_corr}
\end{equation}
These expressions retain the Green-function weighting, wall-normal correlations, and phase cancellations omitted by ${\rm r.m.s.}(\mathcal M)$. The plus-normalised density is defined so that $\mathcal S_{\mathcal M}^+(y^+)\,dy^+$ contributes to $\langle p_{\mathcal M}^{+2}\rangle$.

Figure~\ref{fig:dvdx_linear_source_attribution}(a) shows the resulting $\mathcal S_{\mathcal L}^+(y^+)$. The density rises through the viscous and buffer layers, peaks at near-wall distances, and then decays through the logarithmic and outer regions more rapidly than the $(y^+)^{-1}$ guide. The negative part of the signed density is small and does not form a compensating outer contribution.

This observation is most naturally interpreted as a statement about the possible asymptotic contribution. If, over an inertial range $y_0^+\ll y^+\ll \delta^+$, the density were Reynolds-number invariant in inner units, then a non-cancelling $\mathcal S_{\mathcal L}^+\sim C/y^+$ tail would produce a contribution $C\ln(\delta^+/y_0^+)$ as $\delta^+$ increases. Any steeper decay leads instead to a contribution that approaches a constant as the upper limit extends. The measured $\mathcal S_{\mathcal L}^+$ in figure~\ref{fig:dvdx_linear_source_attribution}(a) is consistent with the latter behaviour.

The key point is therefore that the variance density, not the local amplitude, determines the Reynolds-number dependence. The difference between panels (a) and (b) highlights this: the local product $|2\Omega_z^+|\,{\rm r.m.s.}(\partial_{x^+}v^+)$ does not account for Green-function weighting or cancellation, and does not predict the decay of $\mathcal S_{\mathcal L}^+$.

The conclusion from this section is necessarily limited. The $\delta^+\approx550$ dataset does not establish the asymptotic Reynolds-number decomposition of $\langle p^{+2}\rangle$. It does show that, in the present flow, the exact self-source density of $\langle p_{\mathcal L}^{+2}\rangle$ is near-wall dominated and decays more rapidly than the form required to produce a logarithmic contribution. This is consistent with interpreting $B_{\mathcal L}$ as a predominantly inner-set contribution, while any $\ln\delta^+$ growth of the total variance arises from components whose source densities remain distributed over the expanding inertial layer. The next section considers this possibility for $\mathcal Q$.
    
\section{Fissures as compact carriers of the source terms}\label{sec:fissures}

  \begin{figure}
    \centering
    \includegraphics[width=\linewidth]{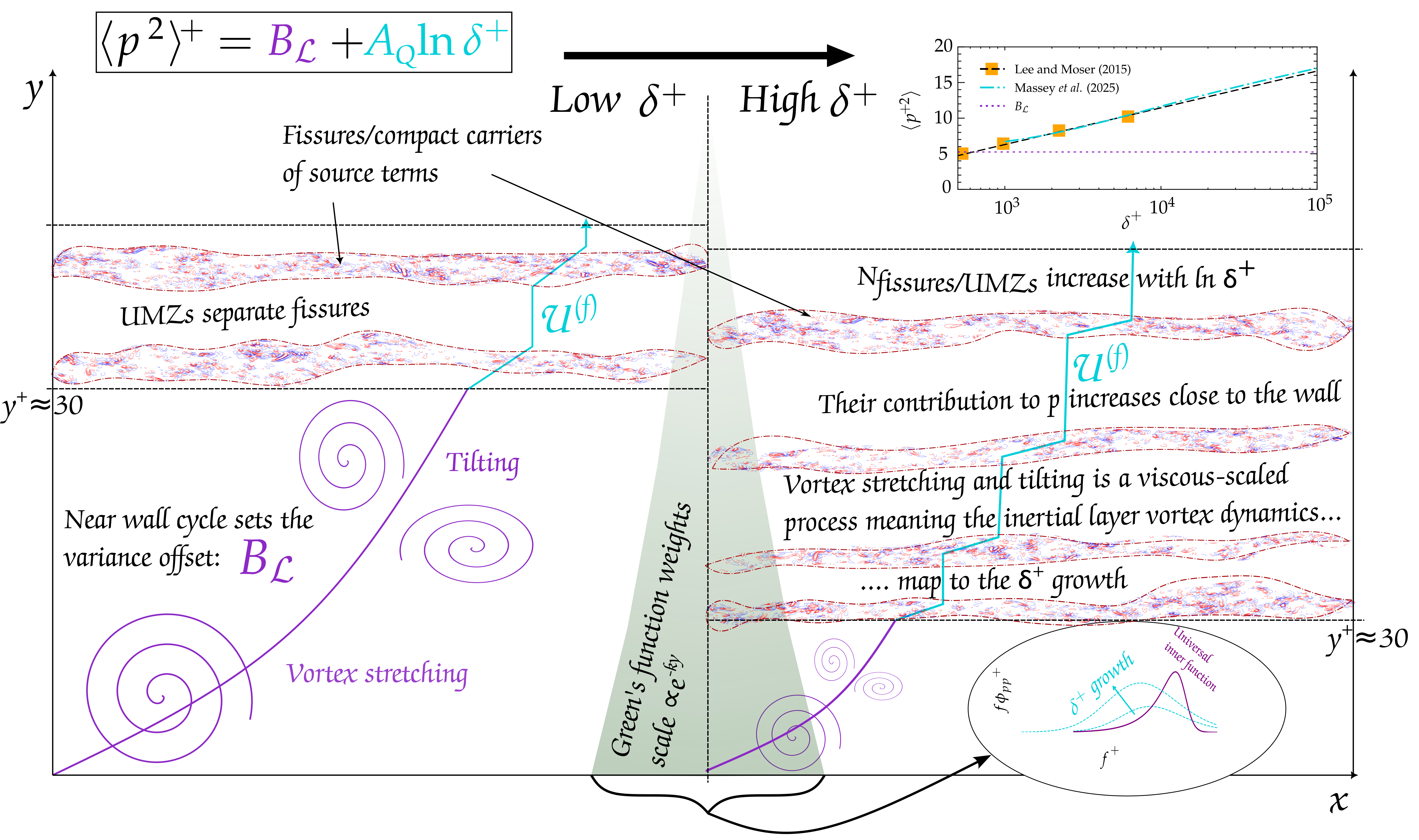}
    \caption{Schematic of the physical process that yields the $\ln\delta^+$ growth of $\langle p^{+2}\rangle$. Vortex stretching/tilting produces thin, sheetlike fissures that carry the Poisson sources compactly across the inertial layer and imprint at the wall through an elliptic Green's mapping that favours low wall-parallel wavenumbers.}
    \label{fig:schematic}
  \end{figure}

  Moving away from the mid-$\delta^+$ dataset, we aim to describe the mechanistic link between the pressure-Poisson sources and the vortical-fissure picture of the inertial layer. The theoretical arguments presented rest on the compact spatial support of the sources within the fissures. We define compact support as regions where the source terms are significantly nonzero and spatially localised. Figure~\ref{fig:schematic} summarises the formation and alignment of these fissures and their variation with $\delta^+$. We consider a single fissure neighbourhood and introduce local coordinates
\begin{equation}
  (s,n,t)=(x,y,z)+O(\theta),\qquad \theta\ll 1,
\label{eq:local fissure coords}
\end{equation}
with $s$ (approximately streamwise), $n$ (approximately wall-normal), and $t$ (approximately spanwise) \citep{cui_geometry_2025}. In the fissure frame, the velocity is decomposed into a fissure-aligned (skeletal) probabilistic mean \citep{bautista_uniform_2019} and residuals,
\begin{subequations}\label{eq:fissure-velocity}
  \begin{align}
    \tilde{u}^{(f)}(s,n,t) \;&=\; \mathcal{U}^{(f)}(n) \;+\; u^{(f)}(s,n,t) \;\approx\; \mathcal{U}^{(f)}(y) \;+\; u^{(f)}(x,y,z),\\
    \tilde{v}^{(f)}(s,n,t) \;&=\; v^{(f)}(s,n,t) \;\approx\; v^{(f)}(x,y,z),\\
    \tilde{w}^{(f)}(s,n,t) \;&=\; w^{(f)}(s,n,t) \;\approx\; w^{(f)}(x,y,z),
  \end{align}
\end{subequations}
where $\mathcal{U}^{(f)}$ is the fissure-aligned most probable (UMZ-like) streamwise profile and $u^{(f)}$, $v^{(f)}$, and $w^{(f)}$ are zero mean under a fissure-aligned conditional average, $\langle (u_i)^{(f)}\rangle_f=0$. The global mean remains $U_i=(U(y),0,0)$. Consistent with the UMZ-vortical-fissure picture, the skeletal profile is a near step with jump $\Delta \mathcal{U}^{(f)}=O(u_\tau)$ across thickness $\delta^{(f)}$, such that
\begin{equation}
  \partial_y \mathcal{U}^{(f)} \;\sim\; \frac{\Delta\mathcal{U}^{(f)}}{\delta^{(f)}} \;\approx\; \frac{u_\tau}{\delta^{(f)}} \quad\text{with }\ \delta^{(f)}\ll \ell_{s,t}.
\end{equation}
This yields the local velocity gradient
\begin{equation}
  \partial_i \tilde{u}_j^{(f)} \;\approx\;
  \underbrace{\begin{pmatrix}
    0 & 0 & 0\\
    \partial_y \mathcal{U}^{(f)} & 0 & 0\\
    0 & 0 & 0
  \end{pmatrix}}_{\text{mean fissure shear}}
  \;+\;\partial_i u_j^{(f)} \, .
  \label{eq:local-grad}
\end{equation}

\rthree{Before turning to the spectral content of the source decomposition that follows, we emphasise that $k$ throughout this section refers to the wall-parallel wavenumber magnitude entering the Green's function in equation \eqref{eq:poisson_integral_walls}. The wall-normal thinness of a fissure (its small $\delta^{(f)}$) does not enter this Fourier decomposition; what matters for wall transmission is the source's footprint in the wall-parallel $(x,z)$ plane. A fissure is thin in $y$ but coherent over along-fissure scales $\ell_{s,t}\gg\delta^{(f)}$, so its wall-parallel spectrum is concentrated at relatively low $k$ even though the source amplitude within the fissure is large and locally sheet-like.}

To expose compact support of the nonlinear source $\mathcal{Q}=-\partial_i\partial_j(u_i u_j)$ in local coordinates, we write the fluctuations as the sum of a skeletal contribution (support confined to the fissure) and residuals that are zero mean in the fissure frame,
\begin{equation}
  u_i \;\approx\; \underbrace{\big(\mathcal{U}^{(f)} - U\big)}_{U^{(sk)}(y)}
  \;+\; \underbrace{(u^{(f)},v^{(f)},w^{(f)})}_{u^\star(s,n,t)}.
  \label{eq:skeletal-velocity-decomp}
\end{equation}
Substituting equation \eqref{eq:skeletal-velocity-decomp} yields
\begin{align}
  \mathcal{Q}
  &\,=\; -\partial_i\partial_j(U^{(sk)}_i U^{(sk)}_j)\;
      -\,2\,\partial_i\partial_j\big(U^{(sk)}_i u^\star_j\big)\;
      -\,\partial_i\partial_j\big(u^\star_i u^\star_j\big). \label{eq:nonlinear-decomp}
\end{align}
Because $U^{(sk)}$ has only an $s$ component depending on $n$, $\partial_i\partial_j(U^{(sk)}_i U^{(sk)}_j)=0$. \rthree{The residual product $u^\star_i u^\star_j$ in the last term is volumetric and turbulent throughout the UMZ interior, with no preferred along-plane coherence length, so its wall-parallel spectrum is broadband with substantial content at high $(k_x,k_z)$; under the elliptic Green's mapping, this high-$k$ content is exponentially attenuated before reaching the wall.} The cross-term simplifies to
\begin{equation}
  -\,2\,\partial_x\partial_y\big(U^{(sk)} v^{(f)}\big)
  = -\,2\,\partial_x\Big[(\partial_y U^{(sk)})\,v^{(f)} + U^{(sk)}\,\partial_y v^{(f)}\Big].
\end{equation}
In the fissure neighbourhood, $\delta^{(f)}\ll\ell_{s,t}$, so $\partial_n\gg \partial_{s,t}$ and the first product, proportional to $\partial_y U^{(sk)}\sim u_\tau/\delta^{(f)}$, dominates. Thus,
\begin{equation}
  {\;\mathcal{Q}^{(f)} \;\approx\; -\,2\,\partial_x\partial_y\Big(U^{(sk)}\,v^{(f)}\Big)\;}
  \label{eq:nonlinear-fissure}
\end{equation}
is compactly supported on the fissure. Integrating across the $O(\delta^{(f)})$ thickness yields the fissure-source density
\begin{equation}
  \int \mathcal{Q}^{(f)}\,dy \;\approx\; -2\partial_x\llbracket U^{(sk)}v^{(f)} \rrbracket,
  \qquad
  \llbracket \cdot \rrbracket \equiv \text{jump across the fissure.}
  \label{eq:nonlinear-fissure-int}
\end{equation}
\rthree{The spectral content of $\mathcal{Q}^{(f)}$ is therefore set by the along-fissure variations of the flux $U^{(sk)} v^{(f)}$, whose scales are $O(\ell_{s,t})\gg\delta^{(f)}$, so the resulting wall-parallel content sits at low $k$.}

For the linear source, $\partial_n U_s\neq 0$ only within the fissure. Thus,
\begin{equation}
  {\;\mathcal{L}^{(f)} \;=\; -\,2\,(\partial_n U)\,\partial_s v^{(f)}
  \;\approx\; -\,2\,(\partial_y U)\,\partial_x v^{(f)}\;}
  \label{eq:linear-fissure}
\end{equation}
is likewise localised by $\partial_n U$, with its spectrum set by the along-fissure gradients $\partial_s v^{(f)}$ (low $k$). In short, both $\mathcal{Q}^{(f)}$ and $\mathcal{L}^{(f)}$ are fissure supported and streamwise biased in wavenumber.

\begin{figure}
  \centering
  \includegraphics[width=\textwidth]{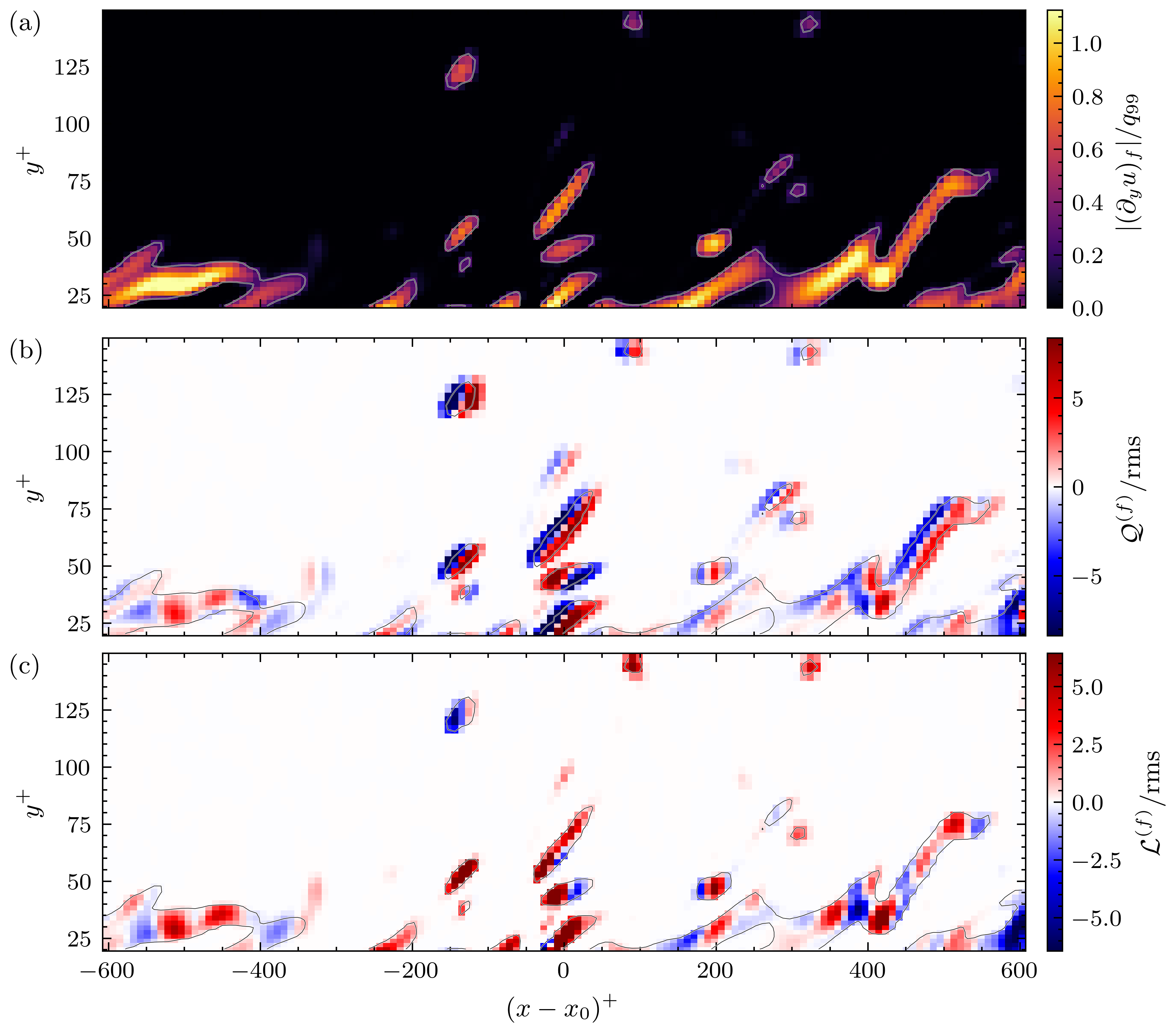}
  \caption{Instantaneous $x$--$y$ snapshot of the fissure-supported source proxies, with $\partial_y u$ used as a direct proxy for the skeletal shear $\partial_y U^{(sk)}$ (see text). (a) Thresholded shear indicator $|(\partial_y u)_f|/q_{99}$, with $(\partial_y u)_f=\chi_f\,\partial_y u$, $\chi_f$ a soft (tanh) threshold defining the fissure support, and $q_{99}$ the $99$th percentile of $|\partial_y u|$ over the analysed wall-normal band (robust colour-bar scale). (b) Nonlinear proxy $\mathcal{Q}^{(f)}=-2\partial_x[(\partial_y u)_f\,v]$ and (c) linear proxy $\mathcal{L}^{(f)}=-2(\partial_y u)_f\,\partial_x v$, each normalised by its r.m.s. Contours in (b,c) repeat the thresholded fissure support.}
  \label{fig:fissure-source-snapshot}
\end{figure}

\rone{Because the skeletal shear $\partial_y U^{(sk)}$ is not directly accessible from instantaneous fields, we adopt the instantaneous wall-normal shear $\partial_y u$ as a direct operational proxy, and define the fissure support via a soft threshold $\chi_f$ applied to $|\partial_y u|$. Substituting $(\partial_y u)_f=\chi_f\,\partial_y u$ for $\partial_n U$ in \eqref{eq:nonlinear-fissure}--\eqref{eq:linear-fissure} yields the source proxies shown in figure~\ref{fig:fissure-source-snapshot}. Across this snapshot, $69\%$ of $|\mathcal{Q}^{(f)}|$ and $91\%$ of $|\mathcal{L}^{(f)}|$ lie within the identified fissure support, confirming that the retained sources are localised on the high-shear skeleton rather than distributed through the surrounding uniform-momentum interiors.}

\rthree{Finally, the wall imprint of these compact sources is dictated by the elliptic Green's function. Because the wall-parallel Green's weight decays approximately as $e^{-k\,y}$, the broadband wall-parallel content of UMZ-interior fluctuations is strongly filtered, whereas the fissure source --- although thin in $y$ --- projects onto \emph{low} wall-parallel wavenumbers because of its large along-fissure extent ($\ell_{s,t}\gg\delta^{(f)}$) and therefore transmits efficiently to the wall.} This picture is consistent with the UMZ-vortical-fissure structure in which $\Delta U=O(u_\tau)$ across fissures of thickness $\delta^{(f)+}\sim(\delta^+)^{1/2}$ \citep{klewicki_description_2013,bautista_uniform_2019}.

\section{Estimate of $A_{\mathcal{Q}}$}\label{sec:AQ_estimate}

  We now use the fissure-supported nonlinear source and the elliptic mapping to obtain an order-of-magnitude estimate for the prefactor $A_{\mathcal Q}$ in \eqref{eq:pressure_variance_law}. From \eqref{eq:nonlinear-fissure-int}, a thin fissure at height $y$ supplies a sheet-integrated nonlinear source. Across the inertial region the streamwise velocity jump satisfies $\Delta U = O(u_\tau)$, and we write $\Delta U^+ \equiv \Delta U/u_\tau$. We denote the inertial-layer r.m.s.\ of $v^+$ by $v_{\rm rms}^+$. Approximating $\partial_x v^{(f)}\sim v_{\rm rms}/\ell_s$, with $\ell_s$ an along-fissure correlation length, the sheet source density scales as
  \begin{equation}
    \int \mathcal{Q}^{(f)} \,{\rm d}y \sim \frac{2 \Delta U\, v_{\rm rms}}{\ell_s} \text{ .}
  \end{equation}

  In the log region, attached-eddy arguments \citep{townsend_structure_1951} and the UMZ--vortical-fissure picture suggest that the relevant wall-parallel scales grow linearly with $y$. Following \citet{wei_properties_2005,klewicki_properties_2021}, we write the typical layer width as $W(y)=y/\phi_c$, with the Fife similarity parameter $\phi_c\approx 1.62$. Then taking
  \begin{equation}
    \ell_s (y) = \alpha W(y) = \alpha \frac{y}{\phi_c} \text{ ,}
  \end{equation}
  where $\alpha=O(1)$ measures the ratio of along-fissure correlation length to UMZ width. The segment-length statistics of large-scale UMZ interfaces in \citet{cui_geometry_2025} show that the wavy segments that compose strongly linear edges have a characteristic length of order $0.1h$ over a wide range of streamwise extents, while the re-analysis of UMZ thicknesses in \citet{cui_geometry_2025} and \citet{heisel_mixing_2020} confirms that $H(y)\sim y$ throughout the logarithmic region. Together, these results support an $O(1)$ ratio $\ell_s/W(y)$ and hence $\alpha=O(1)$. In inner units this gives $y^+/\ell_s^+ = \phi_c/\alpha$, independent of $y^+$ over the inertial layer.

  The wall kernel behaves as $G\sim -k^{-1}{\rm e}^{-k y}$ for $k\gg1$ (see \eqref{eq:Greens k gtr 1}). The attribution maps in figure~\ref{fig:source_spectra} and the envelope in \eqref{eq:elliptic_line} indicate that the dominant contribution to the wall pressure from a layer at $y$ comes from wall-parallel wavenumbers satisfying $k y \approx 1$. Evaluating the kernel on this line gives a factor $y\,{\rm e}^{-1}$. Combining this with the sheet source scaling yields a single-fissure contribution in inner units
  \begin{equation}
    p^{(1)+} \sim C_{\mathcal Q}^{1/2}\Big(2\Delta U^+v_{\rm rms}^+\Big)
    {\rm e}^{-1}\frac{y^+}{\ell_s^+} \text{ ,}
  \end{equation}
  where $C_{\mathcal Q}=O(1)$ collects kernel prefactors, the modest anisotropy of the fissure-aligned source, and weak bandwidth effects associated with the $k y\simeq 1$ envelope. Using $y^+/\ell_s^+=\phi_c/\alpha$ then gives a single-layer variance
  \begin{equation}
    \big\langle (p^{(1)+})^2\big\rangle
    = C_{\mathcal Q}\Big(2\Delta U^+v_{\rm rms}^+\Big)^{2}
    {\rm e}^{-2}\Big(\tfrac{\phi_c}{\alpha}\Big)^{2} \text{ .}
  \end{equation}

  In the UMZ--fissure hierarchy, the characteristic layer positions satisfy $y_{i+1}=\phi_c y_i$, so the number of inertial-layer levels from $y^+\approx \sqrt{\delta^+}$ up to $y^+\approx\delta^+/2$ is
  \begin{equation}
    N(\delta^+) \approx \frac{1}{2\ln\phi_c}\,\ln\delta^+ \text{ .}
  \end{equation}
  Assuming that correlations between distinct layers are weak, the nonlinear contribution to the wall-pressure variance may be written as
  \begin{equation}
    \langle p^{+2}\rangle_{\mathcal Q} \approx B_{\mathcal Q}
    + N(\delta^+)\big\langle (p^{(1)+})^2\big\rangle \text{ ,}
  \end{equation}
  so that
  \begin{equation}
      A_{\mathcal Q}
      = \frac{C_{\mathcal Q}}{2\ln \phi_c}
      \left(2\Delta U^+v_{\rm rms}^+\frac{\phi_c}{\alpha}{\rm e}^{-1}\right)^{2} \text{ .}
    \label{eq:AQ_final}
  \end{equation}

  It is convenient to factor out the purely geometrical numerical prefactor associated with the inertial-layer hierarchy. Writing \eqref{eq:AQ_final} as
  \begin{equation}
    A_{\mathcal Q}
    = C_{\mathcal Q}\,\mathcal M(\phi_c)
    \left(\frac{\Delta U^+v_{\rm rms}^+}{\alpha}\right)^{2} \text{ ,}
  \end{equation}
  we define
  \begin{equation}
    \mathcal M(\phi_c)
    = \frac{2\phi_c^{2}}{\ln\phi_c}\,{\rm e}^{-2} \text{ .}
  \end{equation}
  For the Fife similarity parameter $\phi_c \approx 1.62$, this gives $\mathcal M(\phi_c)\approx 1.47$, so that the remaining sensitivity of $A_{\mathcal Q}$ is contained entirely in the factors $\Delta U^+$, $v_{\rm rms}^+$, $\alpha$ and $C_{\mathcal Q}$.

  Measurements of the velocity jump across UMZ interfaces in high-$Re$ boundary-layer PIV and DNS show $\Delta U^+=O(1)$ and typically slightly greater than unity. The UMZ/VF model of \citet{bautista_uniform_2019} and the mean-vorticity analysis of \citet{klewicki_description_2013} both indicate that $\Delta U$ scales on $u_\tau$ with a range of $\Delta U^+\in[1, 1.6]$ and $\langle \Delta U^+\rangle \gtrsim 1$ \citep{bautista_uniform_2019}, while the interface study of \citet{de_silva_interfaces_2017} finds velocity jumps concentrated in the range $\Delta U^+ \in [1, 2]$. 
  DNS channel flow \citep{hoyas_scaling_2006} up to $\delta^+ \approx 2,000$  likewise show that the inertial-layer wall-normal r.m.s.\ velocity satisfies $v_{\rm rms}^+ = O(1)$, with representative values $v_{\rm rms}^+\sim [0.9, 1.1]$. Taken together, these results motivate using a central estimate $\Delta U^+\approx 1.1$ and $v_{\rm rms}^+\approx 1$ for order-of-magnitude purposes.

  The segment-length and spacing statistics of \citet{cui_geometry_2025} further support an $O(1)$ ratio of along-edge coherence length to UMZ thickness, so for the estimate below we take $\alpha\approx 1$. Finally, $C_{\mathcal Q}$ encapsulates the modest spread of Green's function kernel weights, bandwidth and anisotropy about the idealised $k y\simeq 1$ line; in the absence of any strong bias we likewise treat it as an $O(1)$ factor.

  Using the central values $\Delta U^+\approx 1.1$, $v_{\rm rms}^+\approx 1$, $\alpha\approx 1$ and $C_{\mathcal Q}\approx 1$ in equation~\eqref{eq:AQ_final} yields
  \begin{equation}
    A_{\mathcal Q}\approx 2 \text{ .}
  \end{equation}
  If instead we let $\Delta U^+$ and $v_{\rm rms}^+$ range over the empirical values quoted above, equation~\eqref{eq:AQ_final} gives a broader, but still order-one prediction,
  \begin{equation}
    1.2 \;\lesssim\; A_{\mathcal Q} \;\lesssim\; 4.6 \text{ .}
  \end{equation}
  Thus, the fissure-based picture combined with wall distance scaling robustly produces an $A_{\mathcal Q}$ of order unity, consistent with the logarithmic growth of $\langle p^{+2}\rangle$ inferred from figure~\ref{fig:inner- outer-function schematic}(b).

\section{Discussion and consequences}

  These results link the source terms of the pressure Poisson equations to physical processes that enable us to clarify the $\delta^+$ scaling of $\langle p^{+2}\rangle$. We start by introducing the variance attribution map that elucidates the $\xi=O(1)$ attenuation of structures for $k\gtrsim10$. We then look at the energetic structure of the sources, showing that $\mathcal L$ is concentrated in the buffer layer, while $\mathcal Q$ is most energetic in the inertial layer. $\mathcal Q = \overline{\mathcal Q} + \mathcal T - \mathcal S$ is the difference of two large, strongly correlated quadratic forms; their partial cancellation is central to how $\mathcal{Q}$ maps to the wall. Next, we show that $\mathcal L$ is linked to the near-wall process that subsequently provides an $O(1)$ offset to the observed $\langle p^{+2}\rangle(\delta^+)$ growth. Finally, we show that fissures separating UMZs in the inertial layer act as compact carriers of these source terms. \ra{However, the detailed evidence assembled here comes from a single channel-flow case at $\delta^+\approx 550$, so the manuscript cannot by itself establish how broadly the inferred source split generalises across $\delta^+$ or flow configuration.}

  The conceptual link between the source terms and processes in the turbulent boundary layer allows us to infer scalings for the sources and, via consideration of the Green's function weighting, $\langle p^{+2}\rangle(\delta^+)$. The fissure thickness, when scaled by outer length, decreases approximately as $1/\sqrt{\delta^+}$ while the streamwise velocity jump across a fissure is $O(u_\tau)$. Furthermore, the number of fissures increases approximately as $\ln \delta^+$~\citep{bautista_uniform_2019}, and the fraction of total vorticity contained in the fissures increases with $\delta^+$~\citep{klewicki_description_2013}. The fissure thickness sets the scale for the normal gradients in equations~\eqref{eq:nonlinear-fissure}-\eqref{eq:linear-fissure}, while the velocity jump sets the scale for $U^{(sk)}$. The increasing number of fissures and their increasing proportion of total vorticity content with $\delta^+$ imply that the amplitude and intermittency of $\mathcal{Q}$ increase with $\delta^+$, while $\mathcal{L}$ remains approximately constant as it is tied to near-wall processes. Therefore, considering the observed form of the wall-pressure-variance scaling,
  \begin{equation*}
    \langle p^{+2}\rangle\;=\;B_{\mathcal L}+A_{\mathcal Q}\ln\delta^+ \text{ ,}
  \end{equation*}
  we identify $B_{\mathcal L}$ as the asymptotically $\delta^+$-independent offset from $\mathcal{L}$ and $A_{\mathcal Q}\ln\delta^+$ as the $\delta^+$-dependent contribution from $\mathcal{Q}$. $A_{\mathcal Q}\ln\delta^+$ is therefore tied to the fissure statistics and their evolution with $\delta^+$. \ra{That interpretation should nevertheless be viewed as provisional: with only one Reynolds number analysed directly, the present manuscript supports the plausibility of this decomposition but does not uniquely determine the rate at which the coefficients approach their asymptotic forms, nor does it rule out subleading corrections that could matter over a $\delta^+$ range. Even so, the current dataset is already rich enough to show the source localisation and wall-mapping signatures that motivate the proposed decomposition.} Finally, we provide an order-of-magnitude estimate of $A_{\mathcal Q}$ based on the fissure picture that yields a value consistent with observations. The estimate is sensitive to the choice of parameters, and work to refine these choices would be valuable.

  This picture of the mechanisms driving the source terms augments the \rthree{two component inner- outer-spectral} nature of the premultiplied wall-pressure spectrum identified by \citet{massey_eddy_2025}\rthree{: the explicit link to the source-based view is a subject of ongoing work}. The \rthree{working hypothesis} universal inner function (their $g_1$) arises from the near-wall contribution and the break frequency is defined as the frequency/length scale where the buffer layer contribution is at its maximum. \rthree{The present variance analysis should not be interpreted as a prediction of the Reynolds-number evolution of the inner spectral peak itself; resolving that question would require source-decomposed $k_x$--$k_z$--$\omega$ spectra over $\delta^+$.} The outer function (their $g_2$) arises from the inertial layer contribution with the break frequency associated with length scales on the scale of the boundary layer thickness (consistent with inertial layer structures).  The present framework provides a spatial description of the mechanisms contributing to $\langle p^{+2}\rangle$ and lays the foundations to understand the spatial contribution to the full 2-D wavenumber frequency map of the pressure fluctuations. 

  For control and design, where to act follows from the source split. Targeting the inertial layer to weaken intensity or intermittency reduces $A_{\mathcal Q}$, while targeting the buffer layer primarily modifies $\mathcal L$ and therefore the offset $B_{\mathcal L}$. Because $\xi=O(1)$, actuation localised at $y_a$ imprints most strongly near $k\sim O(1)/y_a$, offering a direct spectral target for feedback or open-loop strategies aimed at the most relevant per-decade inertial layer contribution.

  The present source split is closely related to the classical vorticity/strain form of the Poisson equation proposed by \citet{bradshaw_note_1981} and discussed by \citet{adrian_comment_1982}, who stressed that any such decomposition is algebraically non-unique and that the physically important distinction is between local, irrotational pressure and the nonlocal rotational contribution. Recasting the equation in terms of $\mathcal L$ and $\mathcal Q$ shows that this `rotational' pressure is, in practice, carried by compact vortical fissures in the inertial region, consistent with the vorticity-centric descriptions of \citet{klewicki_description_2013,morrill-winter_influences_2013} and the UMZ-fissure model of \citet{bautista_uniform_2019}. Combined with the elliptic Green's weighting identified in earlier DNS studies \citep{kim_structure_1989,chang_relationship_1999,anantharamu_analysis_2020}, this picture explains how a $\delta^+$-independent near-wall contribution from $\mathcal L$ supplies the offset $B_{\mathcal L}$, while the hierarchy of fissure-supported $\mathcal Q$ sources yields the $A_{\mathcal Q}\ln\delta^+$ growth of $\langle p^{+2}\rangle$, thereby providing a mechanistic underpinning for the matched-asymptotic description of wall-pressure variance \citep{panton_correlation_2017}.

  In summary, this conceptual picture maps the interface between UMZs in turbulent boundary layers to the spectral content of wall pressure and provides a compact lever for models, diagnostics, and control. By tracing how sheetlike structures feed the nonlinear Poisson source and are transmitted to the wall through a radially weighted Green's mapping, this framework connects wall-pressure behaviour to a vorticity-centric description of inertial layer dynamics \citep{klewicki_description_2013} and supplies a mechanistic rationale for the logarithmic growth highlighted by matched-asymptotic theory \citep{panton_correlation_2017}. \ra{Its quantitative generality, however, remains to be tested by repeating the source decomposition and attribution analysis over a wider range of $\delta^+$ and, ideally, across multiple wall-bounded flow geometries. The present $\delta^+\approx 550$ case should therefore be viewed as an illustrative proof of mechanism rather than as a final test of universality.}

\backsection[Acknowledgements]{\rone{The authors thank an anonymous referee for the suggestion that led to figure~\ref{fig:fissure-source-snapshot}, which directly visualises the compact support of the linear and nonlinear sources on the vortical fissure skeleton and meaningfully strengthens the mechanistic story.}}
\backsection[Funding]{The support from DARPA under award HR0011-24-9-0465 and the support of ONR to CTR under grant N000142312833 is gratefully acknowledged.}
\backsection[Declaration of interests]{The authors report no conflict of interest.}
\backsection[Data availability statement]{The data that support the findings of this study will be made available upon request.}
\backsection[Author ORCIDs]{J.M.O. Massey, https://orcid.org/0000-0002-2893-955X; J. C. Klewicki, https://orcid.org/0000-0002-4921-3272; B.J. McKeon, https://orcid.org/0000-0003-4220-1583}

\bibliographystyle{jfm}
\bibliography{refs}

\end{document}